# Fiber-based angular filtering for high-resolution Brillouin spectroscopy in the 20-300 GHz frequency range

A. Rodriguez[*], Priya[*], O. Ortíz, P. Senellart, C. Gomez-Carbonell, A. Lemaître,

M. Esmann[1], N. D. Lanzillotti-Kimura[2]

*Centre de Nanosciences et de Nanotechnologies (C2N), Université Paris-Saclay, CNRS,*

*10 Boulevard Thomas Gobert, 91120 Palaiseau, France*

[*] *Equally contributing authors.*

[1] martin.esmann@c2n.upsaclay.fr

[2] daniel.kimura@c2n.upsaclay.fr

Brillouin spectroscopy emerges as a promising non-invasive tool for nanoscale imaging and sensing. One-dimensional semiconductor superlattice structures are eminently used for selectively enhancing the generation or detection of phonons at few GHz. While commercially available Brillouin spectrometers provide high-resolution spectra, they consist of complex experimental techniques and are not suitable for semiconductor cavities operating at a wide range of optical wavelengths. We develop a pragmatic experimental approach for conventional Brillouin spectroscopy that can integrate a widely tunable excitation-source. Our setup combines a fibered-based angular filtering and a spectral filtering based on a single etalon and a double grating spectrometer. This configuration allows probing confined acoustic phonon modes in the 20-300 GHz frequency range with excellent laser rejection and high spectral resolution. Remarkably, our scheme based on the excitation and collection of the enhanced Brillouin scattering signals through the optical cavity, allows for better angular filtering with decreasing phonon frequency. It can be implemented for the study of cavity optomechanics and stimulated Brillouin scattering over broadband optical and acoustic frequency ranges.

Recently, the acoustic phonons are harnessed, in coherent or incoherent form, for different applications including managing heat transport at the nanoscale, probing elastic properties of thin films and nanostructures, non-invasive imaging of biological membranes and cells, and non-contact inspection of solid-state interfaces in atomically thin heterostructures [1,2]. The urge to study and control the propagation of acoustic phonons has resulted in the emergence of a variety of artificial nanophononic structures presenting confinement of acoustic phonons in the gigahertz to terahertz (GHz-THz) regime [1,3–8]. One-dimensional semiconductor superlattice structures are eminently used for selectively enhancing the generation or detection of phonons at few GHz. The acoustic cavities designed using semiconductor superlattices exhibit discrete acoustic modes for the confinement of phonons. Moreover, the semiconductor optical cavities confining photons can be exploited to simultaneously confine GHz phonons [9–11]. The strong overlap of the optical and acoustical fields in a hybrid optophononic cavity leads to the enhancement of the Brillouin scattering signals [12]. The hybrid optophononic cavities in this acoustic frequency range offer a promising platform for optomechanics [13,14]. The presence of confined acoustic modes can be investigated in the frequency domain using Brillouin spectroscopy [8]. and in the time domain using pump-probe techniques [15,16], Brownian motion schemes [17], or noise spectroscopy [18].

Conventional Raman spectroscopy techniques are used to study optical phonons in the THz range, which are spectrally far from the laser line. While these techniques are usually compatible with excitation sources over a wide range in optical wavelengths, they provide an insufficient stray-light rejection in order to observe Brillouin modes as low as a few tens of GHz =~ 1 cm$^{-1}$. Thus, Brillouin spectroscopy techniques are frequently employed to probe the tailored broadband acoustic spectrum of nanostructures. In the past few decades, Brillouin spectroscopy has taken a giant leap forward in terms of measuring the Brillouin frequency shifts ranging from 0.1 GHz - 1THz with ultra-high resolution of 0.003 cm$^{-1}$ = 0.1 GHz [19–24]. The widely used advanced Brillouin spectrometers providing superior performance are based on the virtually imaged phase array (VIPA) equipped with notch filters and on the scanning multiple pass tandem Fabry-Perot interferometers (TFPI) [24]. These spectrometers are very useful in characterizing the viscoelastic properties of matter with high accuracy and spectral resolution. However, the performance of these techniques strongly relies on an optimization of the optics and detectors to work at a fixed wavelength. Optical scans as a function of wavelength imply intricate optical alignments. Optophononic cavities present optical modes in a wide range of wavelengths. This limits the use of commercial Brillouin spectrometers in probing the confined acoustic modes in hybrid optophononic cavities.

In this paper, we discuss a custom-built accessible and versatile Brillouin spectroscopy scheme to measure longitudinal acoustic phonons in the 20-300 GHz frequency range without optical wavelength restriction, thereby overcoming the limitations of existing Raman and Brillouin spectrometers. The experiments are performed on semiconductor optophononic planar cavities with a typical quality factor of 2000 where the Brillouin scattering signal is enhanced by the mode of the optical cavity [12,25,26]. In order to observe the confined acoustic modes in the Brillouin spectrum with sufficiently high spectral resolution and contrast, we propose a combination of angular and spectral filtering techniques. The angular filtering is implemented with a single-mode fiber to efficiently filter out stray-light from the laser and increase the signal to background ratio. The additional spectral filtering is implemented through tandem of an etalon and double Raman spectrometer. This combination has enabled us to observe the low frequency acoustic modes of a cavity at 20 GHz with a simple etalon, which are otherwise concealed in the excitation laser background.

We have studied two samples (I and II) embedding acoustic cavities to confine phonons at 300 GHz and 18.3 GHz respectively through topological properties [27]. Each sample also consists of an optical microcavity with a cavity mode around 900 nm. This optical cavity is designed to increase the photoacoustic response of the structure [25]. For the 300 GHz sample, the acoustic cavity is embedded in the optical cavity as explained later on. For the 18.3 GHz sample, the same structure that allows to confine phonons also confines photons of the same wavelength [9,11].

The principal goal in Brillouin spectroscopy is the extraction of the low intensity Brillouin peaks with maximum suppression of the stray-light in the collected signal. Figure 1 shows the schematic representation of our home-built Brillouin spectroscopy setup. A collimated laser beam from a tunable continuous wave (cw) Ti:Sa laser (M2 SolsTis) operating at the wavelength in resonance with the fundamental optical cavity mode of the studied sample is used as an excitation source. The excitation laser with <5 MHz linewidth is focused on the sample with a spot diameter of 10 μm using a plano-convex lens *OL* (focal length, f = 13 mm). The same lens is used to collect the scattered signal from the sample. The excitation laser beam passes through a mirror *M* mounted on a translation stage that allows us to vary the incident angle with respect to the surface normal.

The setup is designed in an approximated backscattering geometry with near normal excitation. The measurements are performed in the double optical resonance (DOR) condition to enhance the Brillouin scattering. The DOR condition is achieved by exploiting the in-plane photon dispersion of the optical cavity mode shown in Fig. 1. Under this condition, the excitation laser and the Brillouin scattered signal are coupled to the optical cavity mode. The wavenumber **k** of the optical mode in the cavity can be decomposed as **k** = **k$_z$** + **k$_{//}$** where **k$_{//}$** is the in-plane component and **k$_z$** is the normal component. The resonance is achieved for a given **k$_z$**. The in-plane component **k$_{//}$** increases with the angle of incidence. Therefore, keeping **k$_z$** constant and varying **k$_{//}$** implies a blue-shift of the optical cavity resonance when tuning the angle of incidence away from the surface normal. We optimized the collection of the Brillouin signal resulting from the Stokes process. In this process, the energy of the incoming beam is higher than the energy of the scattered Brillouin signal which means that **k$_{//}$** of the incoming beam is larger than that of the scattered signal. The energy of the scattered signal depends on the frequency of the phonons involved in the process. For an incident in-plane wave vector kept constant, we can observe two configurations. For phonons in the 300 GHz range, the energy shift is large enough that the outgoing signal is normal to the surface. Whereas for phonons in the 20 GHz range, the energy shift is significantly smaller so the outgoing signal has **k$_{//}$** ≠ 0 and is scattered with an angle. We benefit from the DOR to spatially filter the Brillouin signal by selecting the scattered signal for a given **k$_{//}$** with a fiber coupler [25,26].

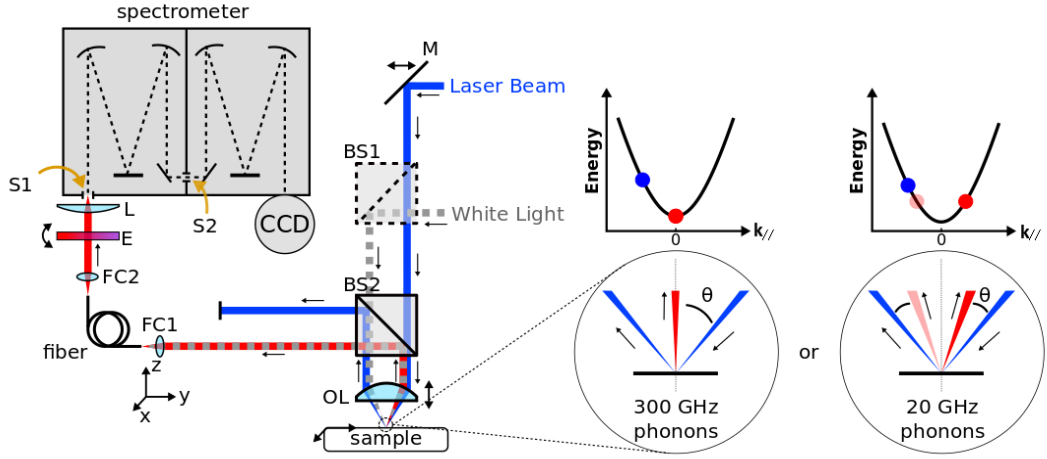

*Fig. 1. Schematic of the Brillouin spectroscopy setup. A tunable cw Ti:Sapph laser excites the sample with an angle of incidence of 13°. The dispersion curves illustrate the DOR condition for different acoustic phonon frequencies providing an angular offset θ (see text for details). The Brillouin signal is collected into a single mode fiber, which allows us to filter out the laser spatially and sent through a tunable etalon into a double grating spectrometer. FC, E, S1, S2 and BS stand for fiber coupler, etalon, spectrometer slits 1 and 2, and beam splitter, respectively.*

Excitation laser as well as scattering signal can be collected through the cavity mode by tuning the angle of incidence and the collection angle ($\theta$), respectively [25]. The angle $\theta$ defines the angular offset with respect to that of the excitation laser in order to achieve the Brillouin scattered signal at the required phonon frequency. The spatial gradient of the sample under study enables us to maximize the coupling of the scattering signal with the optical cavity by changing the excitation laser spot position. The trade-off between the incident angle and sample position is set to obtain the maximal spectral coupling of the excitation and scattered Brillouin signal with the optical cavity mode. In our case for a certain position on the sample, the angle of incidence is fixed at 13° away from the normal for a 3 mm displacement of mirror *M*.

The Brillouin signal is then collected at input port *FC1* consisting of an 11.17 mm focal length fiber coupler connected to a single mode fiber with an NA of 0.13. The single mode fiber allows us to spatially filter the Brillouin signal and reject the reflected excitation laser. The filtering with the fiber is essential to attenuate diffuse laser light that is elastically scattered on the sample and on the multiple optical elements. For phonons at ~20 GHz, we take advantage of the small angular offset $\theta=2°$ to maximize the distance between the reflected laser beam and the Brillouin signal at the input port *FC1*. This is done by collecting the scattered signal on the same side of the normal as the incoming laser beam (see experimental setup in Fig. 1). In this second configuration, the angular filtering thus selects a circular section of NA=0.13*11.7/13=0.12 from the annular Brillouin radiation pattern. For the experimental quality factors and angle of incidence we estimate a maximum Brillouin collection efficiency of 14% inside the fiber. Note that this efficiency is substantially larger for Brillouin emission at normal incidence due to the circular radiation pattern and its larger overlap with the fiber

NA. This signal is then passed through the etalon *E* and sent to the entrance slit *S1* of a double grating spectrometer. Our setup consists of a tandem between a double spectrometer and a simple etalon *E* that permits us to reconstruct a Brillouin spectrum with high spectral resolution compared to the use of a single spectrometer. As the name suggests the double grating spectrometer is a combination of two monochromator chambers connected through an intermediate slit *S2*. The slit *S2* allows us to select the measured optical wavelength range and is particularly used to reduce the level of the stray light generated inside the first chamber from the remaining excitation laser. The signal is then detected by a liquid nitrogen cooled charge-coupled device (CCD) at the exit of the spectrometer. The spectrometer provides an experimental resolution of 9 GHz which is measured as full width at half maximum (FWHM) of the laser source, prior to the use of the etalon. The etalon has a linewidth of 2 GHz and a free spectral range (FSR) of 60 GHz. The etalon is mounted on a motorized rotation stage on a magnetic base and can be easily inserted or removed. We rotate the etalon to vary the angle of incidence and thereby tuning the transmission modes of the etalon. The Brillouin spectrum is a reconstruction from a series of measurements where a mode of the etalon is tuned over a full FSR. The broadband white light source is used to get the position of the transmission lines of the etalon (see Fig. 1). Figure 2(a) shows the design structure of sample I. The acoustic cavity composed of two concatenated distributed Bragg reflectors (DBRs) is enclosed by two optical DBRs forming an optical cavity (see Methods). The measured optical reflectivity spectrum in the inset of Fig. 2(b) shows a dip associated with the optical cavity mode for sample I. The optical cavity mode is centered around 910 nm for an excitation angle of 13°, with FWHM of ~0.4 nm. We have resonantly excited the sample I with an excitation laser tuned at 909.45 nm with an incident power of 16 mW measured before the lens *OL*.

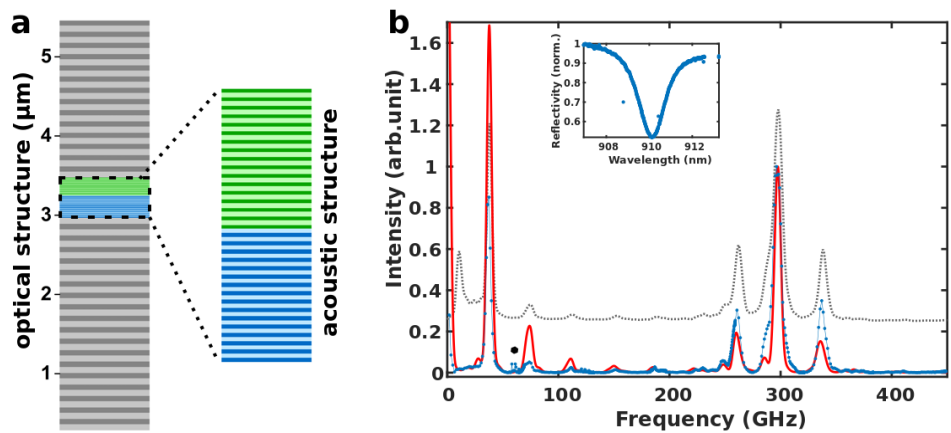

*Fig. 2. (a) Schematic of the optophononic cavity structure of sample I. (b) The calculated (red solid) and experimental (blue dotted line) Brillouin spectrum with etalon measured for sample I. The Brillouin spectrum measured without etalon (grey dashed line) is shown with an offset for comparison. The peak marked with the black dot is parasitic light from the laser. The inset displays the optical cavity mode as a dip in the reflectivity spectrum measured for sample I at an incident angle of 13°.*

Figure 2(b) presents the experimental Brillouin signal prior to (black dashed line) and after (blue dotted line) the insertion of etalon for sample I. The peak at 300 GHz corresponds to the first interface mode between the two acoustic DBRs. The two peaks at ~260 GHz and 335 GHz correspond to modes propagating in the acoustic DBRs [28]. The peaks at lower frequencies < 250 GHz correspond to acoustic modes propagating in the full optical structure. They are multiple harmonics of the mode at 37 GHz. The peak at 37 GHZ is related to the Brillouin mode arising due to the scattering from the GaAs substrate and is also due to modes propagating into the structure. We have compared our experimental results with a calculated Brillouin cross-section using a photoelastic model and transfer matrix method for the structure embedded in GaAs [28]. The calculated spectrum (red solid line) in Fig. 2(b) is convoluted with the linewidth of the etalon to match the resolution of the experimental spectrum. The measured acoustic modes with small oscillations in the experimental spectrum are consistent with the calculations. The peak at 37 GHz also agrees well with our calculations.

Figure 3(a) (dashed line) displays the transmission function of the etalon measured by sending collimated white light through the etalon as shown in Fig. 1. By rotating the etalon, the transmission lines move and span all the frequencies. The angle of rotation is scanned from normal incidence to 1.8° to cover the FSR with a step of 0.02°. The principle of reconstruction of the Brillouin spectrum from the etalon measurements consists of the following steps:

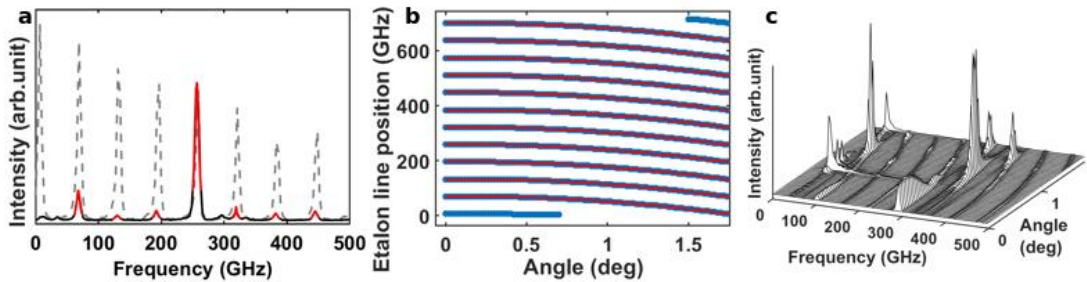

*Fig. 3. Data analysis using an etalon (a) (dashed grey) Transmission line spectrum measured using a broadband white light source at normal incidence. (black) Brillouin signal acquired through the etalon at normal incidence. (red) Lorentzian fit of the Brillouin spectrum. (b) Displacement of the etalon lines as a function of the angle of incidence of the light on the etalon (blue points) measured positions, (red lines) fitted position. (c) Brillouin signal acquired through the etalon as a function of the angle of incidence.*

First, the displacement of transmission lines as a function of the angle of rotation is extracted from the transmission measurement done with the white light; see Fig. 3(b). The displacement of each transmission peak with change in rotation angle represents a parabola. It provides one parabola for each mode of the etalon measured in the spectrometer window in Fig. 3(a). In this situation, the displacement of the transmission line is limited by the frequency resolution of the spectrometer and pixel size of the detector. Therefore, the transmission lines progress in a stepwise fashion. The displacement of each transmission line is fitted with a parabola to smooth the steps and increase the resolution.

Second, the Brillouin scattering signal is acquired in a similar way. Thus, for each Brillouin spectrum acquired through the etalon there is an associated white light spectrum with the position of the transmission lines. Figure 3(a) displays a Brillouin spectrum (black solid line) for the etalon placed at normal incidence on top of the corresponding transmission line (dashed grey line). The Brillouin spectrum is acquired as a function of the angle of rotation of the etalon. The Brillouin peaks appear in the measured spectrum as soon as they coincide with the transmission line of the etalon at that particular rotation angle; see Fig. 3(c). It is important to note that even after blocking the laser using slit S2, the background intensity of the measured spectrum increases when the transmission line coincides with the laser. The peak with the star on top in the measured spectrum shown in Fig. 2(b) is a parasitic line from the laser which arises when the laser is overlapped with the mode of the etalon. That corresponds to a rotation angle of ~0.78°. The acquisition time over a full FSR depends on the acquisition time for one spectrum and is limited by the duration of rotation of the mount. The spectrum is acquired for 4 seconds at each data point. Therefore, if the Brillouin signal is strong enough, the scan can be performed in 10 minutes for a given step-size of 0.02°.

Third, the Brillouin scattering spectra acquired through the etalon are fitted with Lorentzian peaks centered at each transmission line of the etalon (see Fig. 3(a) red lines). It allows us to reduce the light due to the laser reflection on the sample.

Fourth, the Lorentzian fits are integrated on areas with a window of 17 GHz centered on the transmission lines. Each integral is then associated to a position of the transmission line at a given angle. By doing it systematically for each angle, a piece of the Brillouin spectrum is attributed to one transmission line. Therefore, the Brillouin spectrum can be reconstructed by concatenating the pieces. One can note that the strongest Brillouin peaks are not fully rejected in between two transmission lines of the etalon. The method used here could be further optimized by always integrating the signal convoluted with the white light spectrum around the fixed position of the Brillouin peaks.

In order to test the technique at lower frequencies, we studied Sample II which presents a confined acoustic mode at 18.3 GHz. The inset of Fig. 4(a) displays the schematic of the structure where two DBRs are concatenated to simultaneously confine the optical and acoustical fields at the interface. Sample II is also excited resonantly at an excitation laser wavelength of 924.72 nm, with an excitation power of 6.3 mW. The excitation laser is kept with the same angle of incidence of 13°. In this case to satisfy the DOR condition, the scattered signal is no longer collected at normal incidence but with an angle $\theta$ = 2° as shown in Fig. 1, designated for the Brillouin shift of 18.3 GHz which amounts to a difference in optical wavelength of 0.06 nm. Remarkably, the distance between the reflected laser and the signal at the input of the fiber is larger than in the previous case (see Fig. 1), enabling a better fiber-based angular filtering. Here, the mode of the etalon scanned the full FSR using a step of 0.01°.

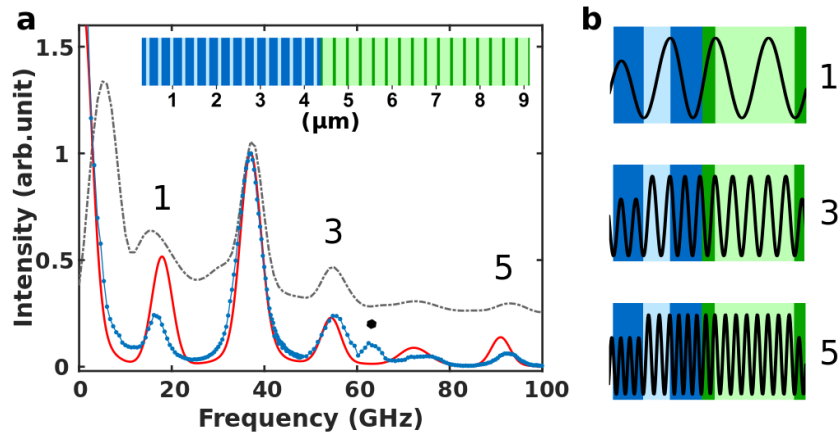

*Fig. 4. (a) The experimental Brillouin spectra without (grey dashed line with an offset) and with (blue dotted line) etalon along with the calculated spectrum (red solid) for sample II. The peak marked with the black dot is the parasitic light from the laser. Schematic of the DBR structure of sample II is displayed as an inset. (b) The acoustic displacement for the 1st, 3rd and 5th harmonic is represented at the interface.*

Figure 4(a) displays the Brillouin spectra measured with and without the etalon together with the calculated spectrum for sample II. The peak at 18.3 GHz is the first interface mode which is observed in DOR condition. The peaks at 56 and 90 GHz correspond also to interface modes between the two DBRs at higher frequencies and are the third and fifth harmonics. The acoustic displacement at the interface is displayed in Fig. 4(b) for the three harmonics observed in our measurement. As before, the peak at 37 GHz is related to the Brillouin mode arising due to the scattering from the GaAs substrate that agrees well with our calculations. The processed spectrum is obtained following the same reconstruction method as sample I. The calculated spectrum obtained for the structure is in complete agreement with the experimental result. The intense peak which appears close to the laser line on the Brillouin spectrum without etalon (grey dashed line) of Fig. 4(a) is due to laser light diffracted by the edge of the slit *S2*. Despite this scattered laser line, the peak corresponding to the interface mode at 18.3 GHz is already visible thanks to the efficiency of the angular filtering. However, its superposition with the remaining unwanted laser background leads to an artificial red shift of the peak. When adding the etalon to perform a better spectral filtering, this background is efficiently suppressed and the signal contrast is significantly improved for the mode at 18.3 GHz.

A similar experimental technique has been discussed in ref [29], which presents a tandem of a gas-pressure controlled Fabry-Perot interferometer and a triple spectrometer to access broadband acoustic frequencies in DBR based microcavities. The technique presents Raman/Brillouin spectra with high resolution of ~0.01cm$^{-1}$ = 0.3 GHz. In contrast to our scheme, the study reports, the tuning of angle of incidence to remain in the DOR condition while collecting the scattered signal at normal incidence. There, a small aperture is used to filter out the Brillouin signal from the reflected laser which results in insufficient stray-light rejection for modes at lower acoustic frequencies with strong parasitic laser lines in the spectrum. In our scheme, we gain better laser rejection at 20 GHz by exciting and collecting the scattered signal away from the normal leading to spatial separation of the reflected laser and of the signal which is filtered by a single mode fiber.

**Conclusion**

In conclusion, we have applied a combination of optical filtering techniques that allow us to access the spontaneous Brillouin scattering signal originated by the confined acoustic mode of a semiconductor microcavity. The technique relies on the angular filtering with a single mode fiber permitted by the angular offset between the incoming laser and scattered signal. It helps to selectively collect the Brillouin signal and to attenuate stray light from the laser. In addition, this technique works over a large range of optical excitation wavelengths using the same set of broadband optics. The etalon has two functions here. On one hand it is used to filter the signal to increase the spectral resolution. On the other it blocks the remains of the laser before the spectrometer to gain high-contrast. Despite the simplicity of the proposed experimental setup, the resolutions (~2 GHz) and stray-light rejection are remarkable in comparison to a Raman spectrometer. With an excitation power of 25 mW, we manage to get well defined peaks within 1 second per datapoint. The ratio signal/background measured at 40 GHz is of the order of 100. The signal/noise ratio and integration time could be further improved by using annular apertures in a different collection approach. Our signals profit from the enhancement of both acoustical and optical fields and provide a prospective tool for the study of stimulated Brillouin scattering at 20 GHz with high-resolution. The simplified detection of high frequency acoustic phonons has implications in the study of phonon-assisted emission from a gain medium such as quantum well or dot embedded in an optophononic cavity. Such high-resolution spectroscopy could also be valuable to probe any change in the density of acoustic modes by probing the phonon sideband of quantum emitters coupled to an acoustic cavity. The proposed experimental scheme thus offers an accessible and versatile platform for exploring cavity optomechanics and phonon lasing at broadband acoustic frequencies.

**Methods**

Both the samples are grown by molecular beam epitaxy on a (001) GaAs substrate. Sample I is made of two optical distributed Bragg reflectors (DBRs) enclosing an optical spacer with an optical path-length of 2.5λ [28] at a resonance optical wavelength around 910 nm. The optical spacer is composed of two concatenated acoustic DBRs with different topological properties designed to confine phonons at 300 GHz. The top (bottom) optical DBR is made of 14(18) periods of 65.1nm/76.3nm $Al_{0.95}Ga_{0.05}As/Al_{0.1}Ga_{0.9}As$. The acoustic top (bottom) DBR is formed by 16 periods (each) of 8.7nm/8.4 nm (7.5 nm/10.0 nm) GaAs/AlAs layers. Sample II consists of two DBRs concatenated to confine simultaneously an acoustic interface mode at 18.3 GHz and an optical interface mode at 920 nm [27,30]. The DBRs are formed by 14 (16) periods of 65.1 nm / 231.1 nm

(195.5 nm / 77.0 nm) GaAs/Al0.95Ga0.05As layers for the top (bottom). The samples were grown with a spatial gradient such that they present position dependent resonance wavelengths.

**Funding**

The authors acknowledge funding from European Research Council Starting Grant No. 715939, Nanophennec. This work was supported by the European Commission in the form of the H2020 FET Proactive project TOCHA (No. 824140). The authors acknowledge funding from the French RENATECH network and through a public grant overseen by the ANR as part of the "Investissements d'Avenir" program (Labex NanoSaclay Grant No. ANR-10-LABX-0035). M.E. acknowledges funding from the Deutsche Forschungsgemeinschaft (DFG, German Research Foundation) Project 401390650.